\documentclass[prl,twocolumn,showpacs,preprintnumbers,amsmath,amssymb,floatfix]{revtex4}
\usepackage{epsfig}
\usepackage{physics}
\usepackage{siunitx}
\usepackage{xr-hyper}
\usepackage{hyperref}

\newcommand{\rem}[1]{}
\newcommand{\Ham}{\widehat{H}}
\newcommand{\btau}{\boldsymbol{\tau}}
\newcommand{\opa}[1]{{\hat{a}^{\phantom \dagger}}_{#1}}
\newcommand{\opadag}[1]{{\hat{a}^{\dagger}}_{#1}}
\newcommand{\BZO}{\scriptscriptstyle{\mathrm{BZ}\setminus {\mathbf 0}}}
\newcommand{\BZ}{\scriptscriptstyle{\mathrm{BZ}}}
\newcommand{\bepsilon}{\boldsymbol{\epsilon}}
\newcommand{\D}{\mathbf{D}}

\newcommand{\G}{\vb{G}}

\renewcommand{\k}{\vb{k}}
\newcommand{\q}{\vb{q}}
\newcommand{\R}{\vb{R}}
\renewcommand{\u}{\vb{u}}
\newcommand{\x}{\vb{x}}
\newcommand{\nep}{\textrm{e}}

\begin{document}

\title{Striped twisted state in the orientational epitaxy on quasicrystals}

\author{Nicola Manini} 
\affiliation{Dipartimento di Fisica, Universit\`a degli Studi di Milano, Via Celoria 16, 20133 Milano, Italy}

\author{Mario Forzanini} 
\affiliation{Dipartimento di Fisica, Universit\`a degli Studi di Milano, Via Celoria 16, 20133 Milano, Italy}

\author{Sebastiano Pagano} 
\affiliation{Dipartimento di Fisica, Universit\`a degli Studi di Milano, Via Celoria 16, 20133 Milano, Italy}

\author{Marco Bellagente} 
\affiliation{Dipartimento di Fisica, Universit\`a degli Studi di Milano, Via Celoria 16, 20133 Milano, Italy}

\author{Martino Colombo}  
\affiliation{Dipartimento di Fisica, Universit\`a degli Studi di Milano, Via Celoria 16, 20133 Milano, Italy}

\author{Dario Bertazioli} 
\affiliation{Dipartimento di Fisica, Universit\`a degli Studi di Milano, Via Celoria 16, 20133 Milano, Italy}

\author{Tommaso Salvalaggio}
\affiliation{Dipartimento di Fisica, Universit\`a degli Studi di Milano, Via Celoria 16, 20133 Milano, Italy}

\author{Andrea Vanossi}
\affiliation{CNR-IOM, Consiglio Nazionale delle Ricerche - Istituto Officina dei Materiali, c/o SISSA Via Bonomea 265, 34136 Trieste, Italy}
\affiliation{International School for Advanced Studies (SISSA), Via Bonomea 265, 34136 Trieste, Italy}

\author{Davide Vanossi}
\affiliation{Department of Chemical and Geological Science, DSCG, University of Modena and Reggio Emilia, Via Campi 103, 41125 Modena, Italy}

\author{Emanuele Panizon}
\affiliation{The Abdus Salam International Center for Theoretical Physics, Strada Costiera 11, 34151 Trieste, Italy}
\affiliation{Area Science Park, Localit\`{a} Padriciano 99, 34149 Trieste, Italy}

\author{Erio Tosatti} 
\affiliation{International School for Advanced Studies (SISSA), Via Bonomea 265, 34136 Trieste, Italy}
\affiliation{The Abdus Salam International Center for Theoretical Physics, Strada Costiera 11, 34151 Trieste, Italy}
\affiliation{CNR-IOM, Consiglio Nazionale delle Ricerche - Istituto Officina dei Materiali, c/o SISSA Via Bonomea 265, 34136 Trieste, Italy}

\author{Giuseppe E. Santoro}
\affiliation{International School for Advanced Studies (SISSA), Via Bonomea 265, 34136 Trieste, Italy}
\affiliation{The Abdus Salam International Center for Theoretical Physics, Strada Costiera 11, 34151 Trieste, Italy}
\affiliation{CNR-IOM, Consiglio Nazionale delle Ricerche - Istituto Officina dei Materiali, c/o SISSA Via Bonomea 265, 34136 Trieste, Italy}

\begin{abstract}
The optimal ``twisted'' geometry of a crystalline layer on a crystal is long known, but that on a quasicrystal is still unknown and open.
We predict analytically that the layer equilibrium configuration will generally exhibit a nonzero misfit angle. 
The theory perfectly agrees with numerical optimization of a colloid monolayer on a quasiperiodic decagonal optical lattice. Strikingly different from crystal-on-crystal epitaxy, the structure of the novel emerging twisted state exhibits an unexpected stripe pattern.
Its high anisotropy should reflect on the tribomechanical properties of this unconventional interface.
\end{abstract}

\date{\today}
\maketitle

The optimal structure which two-dimensional (2D) crystalline monolayers adopt upon deposition onto a lattice-mismatched crystal surface, a frequent occurrence for many current 2D materials and devices \cite{Jong22, Lisi22, Dindorkar23}, is known as a source of counterintuitive phenomena.
Novaco and McTague (NM) \cite{Novaco77,McTague79}, followed by others \cite{Shiba79,Shiba80}, 
predicted
%in the 1970s 
that the two 2D lattices would generally not line up, and instead rotate
by some small twist angle $\theta$ -  a prediction that was promptly verified by Ar deposition on graphite \cite{Shaw78}. 
A variety of 2D lattices, recently including colloidal monolayers \cite{MandelliPRL15,MandelliPRB15}, actually adopt this slightly twisted epitaxy, 
where the gain in corrugation/adhesion energy outweighs the elastic-energy cost.
In the rotated geometry, the misfit dislocations turn from compressive to shear, the latter less expensive owing to the lower shear stiffness \cite{McTague79,MandelliPRB15}.

\begin{figure*}
\centering
\includegraphics[width=\textwidth,clip=]{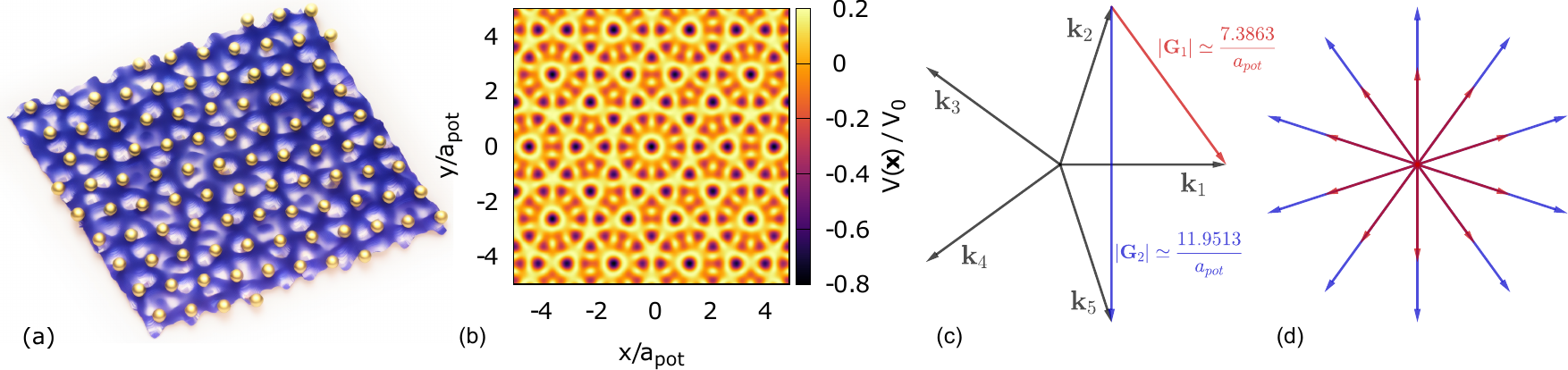}
\caption{\label{combined:fig}
(a) Sketch of a 2D crystalline colloidal layer interacting with a decagonal quasiperiodic energy profile
$V(\vb{x})$ 
generated by light interference.
(b) A portion of the 5-fold symmetry corrugation energy $V(\vb{x})$, which we study here, pictured as a function of position.
(c) The reciprocal lattice vectors $\vb{k}_m$ of potential
$V(\vb{x})$ in Eq.~\eqref{pot:eq}, and the geometric construction of the
$\vb{G}$ vectors for $p=5$. 
(d) The resulting $p(p-1)=20$ $\vb{G}$ vectors.
}
\end{figure*}

Less clear is what should happen to the overlayer orientational epitaxy when the crystalline substrate is replaced by a quasicrystal.
The quasicrystal, even if not translationally periodic,  has well-defined directional axes (Fig.~\ref{combined:fig}a-b), with which the overlayer crystal axes could align or not.
The crystal-on-quasicrystal misfit superstructures are quite different in nature and energetics from the crystal-on-crystal case, and both experiments and theory show that quasicrystal surfaces induce a considerably richer structural complexity in the overlayer \cite{Schmiedeberg08,Mikhael10,Zaidouny14}.
That complexity includes pseudomorphic phases with both crystalline and quasicrystalline conformations (Archimedean-like tiling arrangements) \cite{Mikhael08} and the proliferation of anomalous structures \cite{Mikhael10}, both 
observed in elegant colloidal monolayer experiments.
While neither theory nor existing data so far suggested a misaligned geometry, we predict in this Letter that twisted epitaxy will often occur in the quasicrystalline case, too. Obtained analytically and confirmed by a model 
simulation, the rotation is accompanied in this case by
a striped pattern, quite different from the
the ordinary moir\'e pattern of the crystalline case.

{\em The model -}
Consider a crystalline harmonic 2D monolayer under the influence of a
quasi-crystalline potential $V(\vb{x})$, %with Hamiltonian
\begin{equation}\label{ham:eq}
    \Ham =
    \Ham_0+\Ham_1 =
    \Ham_0 + \sum_j V(\R_j + \hat{\u}_j)
\end{equation}
where 
\begin{equation}
\Ham_0=\sum_{\q}^{\BZO} \sum_s \hbar\omega_{\q,s} \opadag{\q,s} \opa{\q,s}
\end{equation}
is the harmonic intralayer interaction, 
$\opadag{\q,s}$ is the creation operator of a phonon with wavevector $\q$ and polarization $s$, 
\begin{equation} \label{eqn:u_j}
\hat{\u}_j = \frac{1}{\sqrt{N}} \sum_{\q}^{\BZO} \sum_s 
\sqrt{ \frac{\hbar}{2M\omega_{\q,s}} }
\nep^{i\q \cdot \R_j} \bepsilon_{\q,s} \Big( \opadag{-\q,s} + \opa{\q,s} \Big) \,,
\end{equation}
is the displacement of the $j$-th particle from its lattice equilibrium position $\R_j$. 
The phonon eigenvectors $\bepsilon_{\q,s}$ 
($s=\mathrm{L,T}$) 
\footnote{For a monoatomic crystal \cite{Ashcroft}, the matrix $\D(\q)$ is real
for any $\q$: therefore, the eigenvectors $\bepsilon_{\q,s}$ can also be taken real.}
are such that $\bepsilon_{\q,s} \cdot \bepsilon_{\q,s'} =\delta_{s,s'}$
with $\bepsilon_{-\q,s} = \bepsilon_{\q,s}$.
For specificity, here we focus on a hexagonal lattice with equilibrium spacing $a_\text{coll}$ and 
nearest-neighbor elastic coupling $K$.
We assume moreover a rigid quasi-periodic ``substrate'' potential
\begin{align} \label{pot:eq}
V(\x)
= -\frac{V_0}{p^2} \sum_{\G} \nep^{-i\G \cdot \x}
\,,
\end{align}
where the $p(p-1)$ vectors $\G$ are defined as differences of 
$\k_m$ vectors in Sect.~S1 of the Supplemental Material (SM) \cite{SMNovQuasCrys:note}, and shown in Fig.~\ref{combined:fig}.
The potential \eqref{pot:eq} can model the in-plane atomic corrugations that adatoms experience when deposited on
an atomically flat quasicrystalline surface, or the light intensity pattern resulting from the
superposition of $p$ identical coherent laser beams  incoming
from regularly-spaced directions, and with in-plane wavelength
$a_\text{pot}$ \cite{Vanossi12PNAS,Brazda18,Mikhael11,Bohlein12,Bohlein12PRL,Brunner02, Mangold03, Bleil06, Mikhael10}.
Specifically
a quasi-crystalline potential with decagonal symmetry is
generated when $p=5$.

The quasicrystalline potential Eq.~\eqref{pot:eq} produces a
perturbation to the harmonic $\Ham_0$:
\begin{equation}
  \Ham_1 = -\frac{V_0}{p^2} \sum_j \sum_{\G}
  \nep^{-i\G \cdot \R_j} \nep^{-i\G \cdot \hat{\u}_j } \,.
\end{equation}
The dimensionless ratio $g=V_0/(K a^2_{\text{coll}})$
identifies the coupling regime of the model \cite{Floria96}.
When $g \gg 1$, the coupling is strong and it is energetically 
costly for particles to move
away from the the minima of $V(\vb{x})$, while it is comparatively cheaper to
deform the crystalline arrangement.
As a result, particles tend to
remain 
close to the bottom of the potential wells: this regime was addressed by previous work \cite{Mikhael11,Reichhardt11}.
On the other hand, for
$g \ll 1$ the substrate potential $V(\vb{x})$ is a weak perturbation only causing small
deviations from the perfect crystalline arrangement of the overlayer.
The original NM theory \cite{Novaco77,McTague79} described crystal-on-crystal epitaxy in the weak-coupling regime.
Here we extend it to a crystal-on-quasicrystal case in the same regime.

{\em Variational approach - }
We assume that the phonons are in a coherent state \cite{Mahan} of the form:
\begin{align}
  |\Psi \rangle = 
  \nep^{-\frac{1}{2} \sum_{\q,s} |z_{\q,s}|^2} \nep^{ \sum_{\q,s} z_{\q,s} \opadag{\q,s} } | 0 \rangle \,,
\end{align}
where the $z_{\q,s}$ are dimensionless variational parameters, with
$z_{-\q,s}=z_{\q,s}^*$, and 
we evaluate the average energy of this state, 
$E= \langle \Psi | \Ham_0 + \Ham_1 | \Psi \rangle$.

\begin{figure}
\includegraphics[width=0.9\columnwidth,angle=0,clip=]{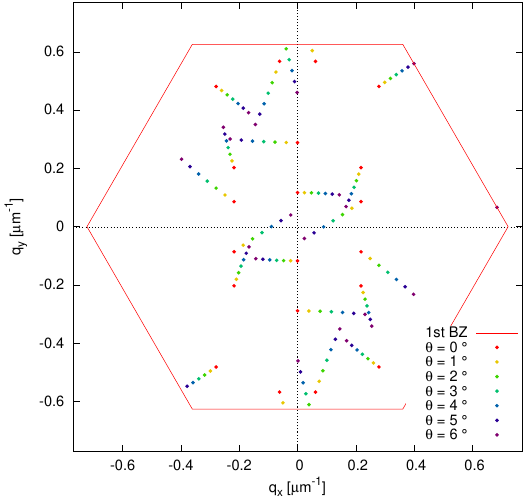} 
\caption{\label{fig:qs}
Red hexagon: the first Brillouin zone of the hexagonal lattice with
$a_{\text{coll}} = \SI{5.8}{\micro\meter}$.
Dots: successive locations of the $\q$ points obtained by Eq.~\eqref{eq:qs}, as the $p(p-1)=20$ vectors $\G$ (drawn in Fig.~\ref{combined:fig}d) of the decagonal quasiperiodic potential ($a_{\text{pot}} = \SI{5.4}{\micro\meter}$) are rotated counterclockwise in $\theta$ steps.
}
\end{figure}

The average of the individual terms is straightforward
to calculate in the coherent state, by
using the fact that $|\Psi\rangle$ is an eigenstate of the phonon
destruction operator \cite{Mahan}, $\opa{\q,s}|\Psi\rangle = z_{\q,s} | \Psi\rangle$. Details are given in the SM \cite{SMNovQuasCrys:note}.
The variational total energy per particle is:
\begin{align} \label{totalE:eq}
\mathcal{E} = \frac{E}{N} &= \sum_{\q}^{\BZO} \sum_s \hbar\omega_{\q,s} | \xi_{\q,s} |^2 \nonumber \\
&\hspace{2mm}
- \frac{V_0}{p^2} \sum_{\G} \nep^{-W_{\G}} \bigg( \frac{1}{N} \sum_j \nep^{-i \G \cdot \R_j} \nep^{-i \G \cdot \u_j} \bigg) \,,
\end{align}
where $\xi_{\q,s}=z_{\q,s}/\sqrt{N}$ guarantees correct scaling in the thermodynamic limit.
Moreover, $\u_j = \langle \Psi | \hat{\u}_j | \Psi \rangle$ is the average displacement field,
a linear function of $\{\xi_{\q,s}\}$. The Debye-Waller factor $W_{\G}$ will be ignored, as 
negligible in a colloidal system.
$\mathcal{E}$ should be minimized with respect to the variational parameters $\{\xi_{\q,s}\}$, whence %Imposing 
$\partial \mathcal{E}/\partial \xi^*_{\q,s}=0$ leads to (see SM \cite{SMNovQuasCrys:note}):
\begin{equation} \label{variational:eq}
\hbar\omega_{\q,s} \xi_{\q,s} =
i \frac{V_0}{p^2} \sqrt{ \frac{\hbar}{2M\omega_{\q,s}} }
 \sum_{\G} (\G\cdot \bepsilon_{\q,s}) 
 F_{\q,\G} \;,
\end{equation}
where 
$F_{\q,\G} = 
  \frac{1}{N} \sum_j \nep^{-i (\q-\G) \cdot \R_j} \nep^{i\G \cdot \u_j}$
depends non-linearly on the $\{ \xi_{\q,s} \}$ through 
$\nep^{i\G \cdot \u_j}$.
In the one-phonon approximation
$\nep^{i\G \cdot \u_j}\to 1+i\G\cdot \u_j$,
momentum conservation 
leads to
\begin{equation} \label{eq:qs}
  \q = \G - \btau 
\end{equation}
where $\btau$ are the reciprocal lattice vectors of the harmonic lattice $\{\R_j\}$.
The optimal energy
takes the form (see SM \cite{SMNovQuasCrys:note}): 
\begin{align} \label{NMgeneralized:eq}
\mathcal{E}_\text{1-ph}
\! &\stackrel{\scriptscriptstyle{N\to \infty}}{=}
 - \frac{V_0^2}{2Mp^4} \sum_{s}^{\text{L,T}}
 \sum_{\G,\btau}\, 
\frac{|\G \cdot \bepsilon_{\q,s}|^2}{\omega_{\q,s}^2}\bigg|_{\q=\G-\btau}
^{\q\in \BZ} 
\,.
\end{align}
Only the $\q-$points satisfying momentum conservation contribute.
For each $\G$, exactly one $\btau$ places $\q$ in the first Brillouin zone (BZ): as a result the number of $\q$'s contributing is the same as
the number of $\G$ vectors, namely 20 for our decagonal potential.
Correspondingly, the equilibrium displacement field $\u_j$ is 
\begin{equation} \label{displ:eq}
   \u_j = -\frac{V_0}{Mp^2} \sum_{s}^{\text{L,T}}
 \sum_{\G,\btau}\, 
\frac{\G \cdot \bepsilon_{\q,s}}{\omega_{\q,s}^2} \bepsilon_{\q,s} \sin(\q\cdot \R_j) 
\bigg|_{\q=\G-\btau}^{\q\in \BZ} 
\,.
\end{equation}

One can further minimize $\mathcal{E}_\text{1-ph}$, to find the optimal global arrangement, by varying
the orientation angle $\theta$ 
of the $\btau$-reciprocal lattice relative to the Fourier $\G$ points of
the quasi-crystalline potential, see SM \cite{SMNovQuasCrys:note} Figs.~S2 and S3.
The frequency denominator in Eq.~\eqref{NMgeneralized:eq} tends to favor small-$\q$ {\em transverse}
modes with lower sound velocity, the optimal arrangement is obtained when the projections 
$|\G \cdot \bepsilon_{\q,s}|$ are largest, and their frequencies the lowest: this observation identifies the shortest $\q$ vector $\q_\text{min}$ as the main contributor for the distortion.
In this respect, the main difference between the hexagonal-on-hexagonal
lattice of NM \cite{Novaco77,McTague79} and the hexagonal-on-decagonal
quasicrystal is that in the former (see SM \cite{SMNovQuasCrys:note} Fig.~S4a) the $\q$ vectors retain the original hexagonal symmetry, while in the latter the only symmetry of the
$\q$ vectors is inversion, i.e.\ they come in opposite pairs, see Fig.~\ref{fig:qs}.

{\em Numerical -}
We implement the finite-size classical counterpart of Eq.~\eqref{ham:eq} in LAMMPS \cite{lammps},
a 2D hexagonal monolayer of point particles interacting via harmonic springs of stiffness $K$ and spacing $a_\text{coll}$. 
Particles move in 
the quasi-periodic substrate potential of Eq.~\eqref{pot:eq}.
Energies are expressed in units of $K a_{\text{coll}}^2$, so that the dimensionless results apply equally to 2D colloids in an optical lattice or to
an atomic overlayer on a quasicrystalline surface.
We simulate a circular sample, with 
particles in the outermost ring fixed at perfect-lattice positions, to mitigate boundary effects and to control the twist angle in the nontrivial range $\SI{0}{\degree}\leq \theta \leq\SI{6}{\degree}$ dictated by symmetry.

\begin{figure}
  \centering
\includegraphics[width=0.99\columnwidth,angle=0,clip=]{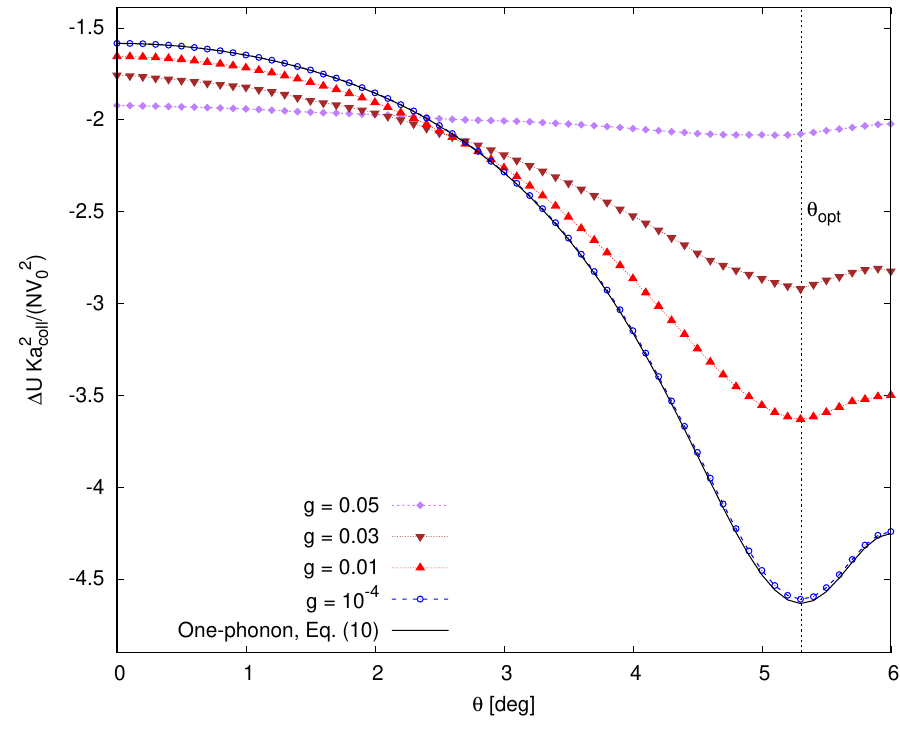}
\caption{\label{fig:novaco-5.4}
Comparison of the numerically relaxed total potential energy per
particle (points), with the
one-phonon result,
Eq.~\eqref{NMgeneralized:eq} (line),  as a function of the twist angle $\theta$.
The geometry is defined by the ratio $a_\text{pot}/a_\text{coll} = \SI{5.4}{\micro\meter} / (\SI{5.8}{\micro\meter})$.
Four different coupling amplitudes from
$V_0 = 10^{-4} \, Ka^2_\text{coll}$ (blue dots) to $V_0 = 0.05 \, Ka^2_\text{coll}$ (violet diamonds) are reported.
The circular sample has $N = 999854$ particles, $N_{\text{tot}} = 1003633$ including the fixed external ring.
Vertical dashed line: the optimal angle $\theta_\text{opt} \simeq \SI{5.31}{\degree}$.
}
\end{figure}

With a  generic ``lattice-incommensurate'' situation  in mind,
we adopt a length ratio $a_\text{pot}/a_\text{coll} = \SI{5.4}{\micro\meter} / (\SI{5.8}{\micro\meter})$ such that
$\G \neq \btau$, hence $\q \neq \vb{0}$, for all twist angles $\theta$. 
The energy lowering $\Delta U/N$ obtained by the numerical relaxation of a circular sample and the one-phonon energy of Eq.~\eqref{NMgeneralized:eq} are compared as a function of twist angle $\theta$ in Fig.~\ref{fig:novaco-5.4}.
In the weak-coupling regime $g\ll 1$ where the one-phonon approximation applies,
the agreement is excellent without any fitting parameters. 
While for
$g\simeq 10^{-4}$ the one-phonon formula predicts the energy lowering with quantitative accuracy, even when the coupling strength is raised to
much larger $g\simeq 10^{-2}$, the overall dependence of the energy on the twist angle $\theta$ remains qualitatively similar to the prediction of Eq.~\eqref{NMgeneralized:eq},
and also the optimal twist angle $\theta_\text{opt} \simeq \SI{5.31}{\degree}$ (for the adopted length ratio $a_\text{pot}/a_\text{coll}$) does not deviate significantly.

\begin{figure*}
\includegraphics[width=0.65\textwidth,angle=0,clip=]{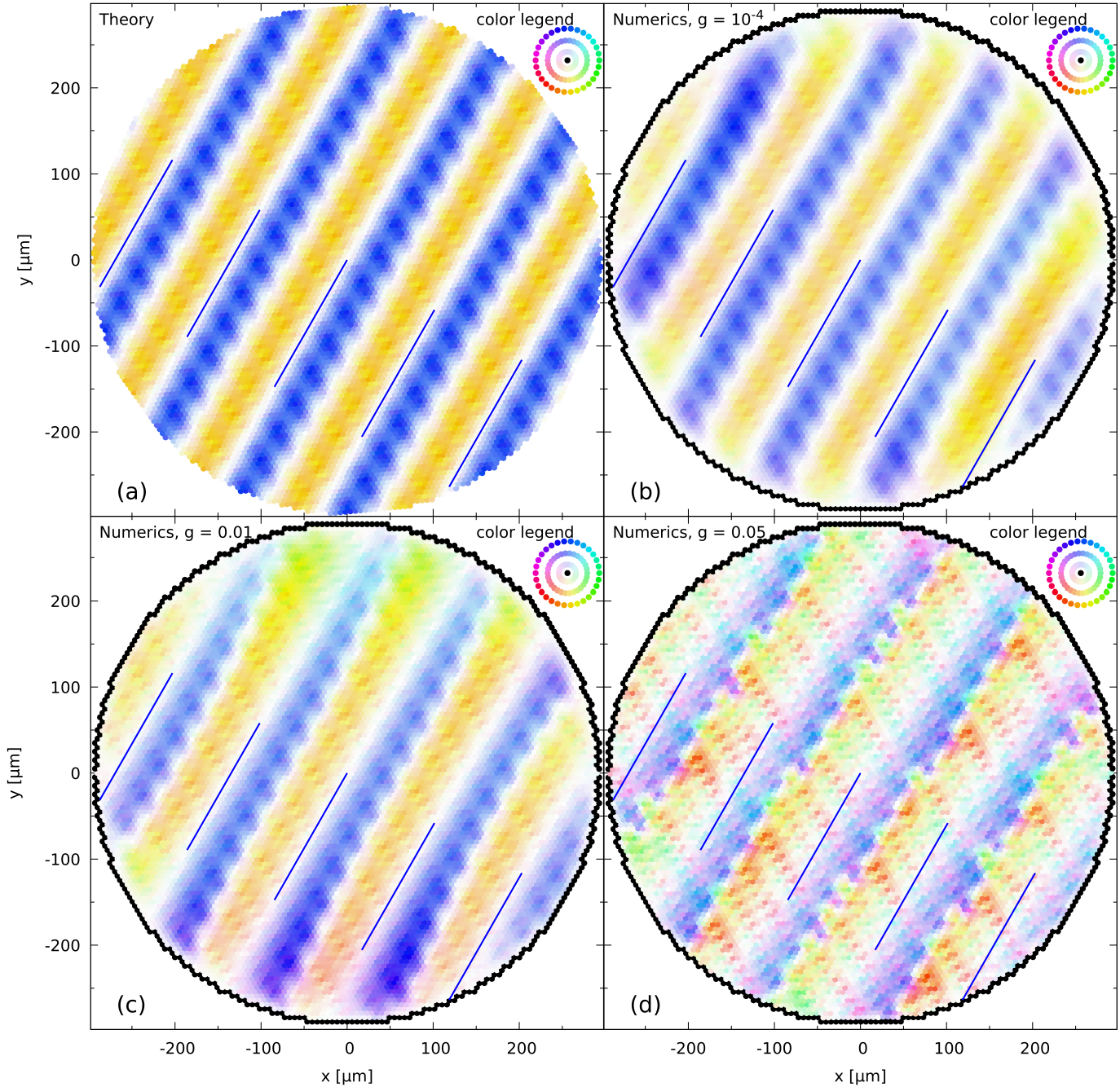}
  \caption{\label{fig:moire}
Comparison of the displacement pattern (a) predicted by Eq.~\eqref{displ:eq} with that of the numerically-relaxed state
of a circular sample of $N=9407$ particles for (b) $g=10^{-4}$, (c) $g=0.01$, (d) $g=0.05$.
Colored dots: particles colored according to their displacement $\vb{u}_j$ from the initial perfect-lattice position;
the color intensity reflects the amount of displacement, the color hue indicates its direction.
Black dots: fixed edge particles, see SM \cite{SMNovQuasCrys:note}.
Blue lines: wavefronts for the shortest $\q$ satisfying Eq.~\eqref{eq:qs}.
Structural parameters:
$a_{\text{pot}} = \SI{5.4}{\micro\meter}$, 
$a_{\text{col}} = \SI{5.8}{\micro\meter}$, 
$\theta = \SI{5.3}{\degree}\simeq \theta_\text{opt}$.
}
\end{figure*}

For each twist angle, 
Eq.~\eqref{displ:eq} predicts
the displacement of each particle from its initial position.
The largest contribution will typically be associated with the softest among the
phonons satisfying Eq.~\eqref{eq:qs}, namely the pair of long-wavelength transverse phonons associated with the shortest
$\q = \pm \q_\text{min}$.
We can verify this observation directly by inspecting the distortions of
the relaxed configuration.
Figure~\ref{fig:moire} compares the displacement pattern predicted by Eq.~\eqref{displ:eq} with that obtained in the relaxed configuration at the same $\theta=\theta_\text{opt}$:
each particle $j$ is colored in a way that represents its
displacement $\vb{u}_j$ away from the perfect unrelaxed crystal position $\vb{R}_j$.
For small $g$ (Fig.~\ref{fig:moire}b),
the displacement patterns match nicely, except for predictable 
deviations
near the sample edge
where
the rigid ring produces just a smooth quenching of the distortion, with little effect on the bulk.
The blue lines drawn perpendicular to
$\q_\text{min}$ and spaced by
$\lambda=2\pi/|\q_\text{min}|$ highlight the pair of phonon modes most
strongly involved in the displacement pattern.
We see that, as a result of the distortion being dominated by just one pair of shortest $\q$ vectors, this hexagonal-on-decagonal interface yields a moir\'e distortion pattern dominated by stripes, as opposed to the hexagonal-on-hexagonal interface leading to a hexagonal moir\'e pattern, widely studied in recent years, 
e.g., in the context of 2D materials \cite{Jong22, Lisi22, Dindorkar23}.
Stronger coupling, exemplified by Fig.~\ref{fig:moire}c,d, leads to a more intricate pattern, involving extra phonon modes.

{\em Discussion -} Summarizing, we predict that the equilibrium epitaxy of a 2D lattice deposited onto a quasi-periodic substrate is described by Eqs.~\eqref{NMgeneralized:eq} and \eqref{displ:eq}.
Obtained analytically and tested numerically, the optimal twist angle prediction generates a definite %non-symmetry 
non-symmetric alignment of the distortion pattern and has the same 
weak-corrugation applicability as the well-established NM 
hexagonal-on-hexagonal
epitaxy.
In comparison, the equilibrium %twisted 
moir\'e pattern differs very importantly.
Dominated by a pair of shortest
reciprocal vectors $\mathbf q_\text{min}$, 
rather than 6 ones placed at the vertices of a regular hexagon, 
the distortion pattern consists of parallel stripes, with lower symmetry and higher anisotropy than for the 
hexagonal-on-hexagonal epitaxy \footnote{Patterns consisting of stripes are expected for any interface between objects of incompatible symmetry, e.g. hexagonal-on-square or square-on-decagonal interfaces \cite{Panizon23}.}.
Such nontrivial angular alignment would be impossible to predict without our quantitative approach.
Experimentally, through e.g.\  properly designed setups with colloidal monolayers on optical lattices \cite{Mikhael08, Mikhael10,Mikhael11}, one should be able to verify these predictions.
The occurrence of a NM lattice twist between crystalline contacting materials strongly modifies
the moir\'e structure, upon which many electronic/vibrational/chemical properties and exotic strongly-correlated and topological phenomena depend \cite{Mora19,Andrei21}.
Here, the emerging highly directional striped moir\'e pattern should reflect in drastic anisotropic features, 
such as the direction-dependent tribological properties \cite{Koren16a} of sliding heterointerfaces realized, e.g.,  by 
crystalline metallic clusters \cite{Guerra10} or
by flakes of 2D materials \cite{Omambac19} on quasicrystalline solid surfaces \cite{Shechtman84,Yadav18}.
In the weak-coupling regime we predict no static friction (superlubricity, see SM Sect.~S2\,A \cite{SMNovQuasCrys:note}) but the kinetic friction upon sliding should involve primarily excitations of the $\mathbf q_\text{min}$ phonon, therefore becoming strongly anisotropic.

\begin{acknowledgments}
Support is acknowledged  from  grant PRIN2017 UTFROM of the Italian Ministry
of University and Research (MUR), and from project NEST funded under the National Recovery and Resilience Plan (NRRP), Mission 4 Component 2 Investment 1.3 - call No.\ 1561 of 11.10.2022 of MUR, funded by the European Union – NextGenerationEU Project code PE0000021, Concession Decree No.\ 1561 of 11.10.2022 by MUR with CUP:D43C22003090001.
GES acknowledges support from PNRR MUR project PE0000023-NQSTI, from PRIN 2022H77XB7 of MUR, 
and from the project QuantERA II Programme STAQS project 
funded by
the European Union’s Horizon 2020 research and innovation programme under Grant Agreement No 101017733.
Work by ET and AV was partly supported  by ERC ULTRADISS Horizon 2020 Contract No.\ 834402.
EP was partially supported by the FVG Regional project with CUP:F53C22001780002.

\end{acknowledgments}

%\bibliography{biblio}

\end{document}